\documentclass[usenatbib]{mn2e}
\usepackage{bm,graphicx,graphics}
\begin{document}

\title{
On bending angles by gravitational lenses in motion}

\author[S. Frittelli]
       {Simonetta Frittelli\\
    Department of Physics, Duquesne University,
        Pittsburgh, PA 15282}
\maketitle
\begin{abstract}

The bending of lightrays by the gravitational field of a ``lens'' that is
moving relative to the observer is calculated within the approximation of weak
fields, small angles and thin lenses. Up to first order in $v/c$ -- and
assuming the acceleration to be much smaller than $v/c$ -- the bending angle,
time delay and redshift of the images are found to be affected by the component
of the speed of the deflector along the line of sight. The correction takes the
form of an overall factor of $1+v/c$ accompanying the mass of the deflector,
leading to an indeterminacy of the order of $v/c$ in the mass of the lens
inferred on the basis of the separation of multiple images. The consequent
correction to the microlensing lightcurve is pointed out, as well as scenarios
where the correction is potentially relevant.

\end{abstract}

%-----------------------------------------------------------------
\section{Introduction}\label{sec:1}

In order to derive a formula for the bending angle of light rays in the
presence of a gravitational deflector, in gravitational lensing it's
commonly assumed that the observer, deflector and source are at rest in
some global coordinate system (see for instance \citet{EFS}).  However,
most gravitational deflectors will not be at rest in the observer's frame
of reference. Typical gravitational lenses are of two kinds: galaxies and
clusters of galaxies at cosmological distances, or massive compact
objects within the Milky Way galaxy. Cosmological deflectors have
peculiar velocities on top of the Hubble flow, and objects in the Milky
Way usually move by the line of sight rather quickly.  Still, the effect
of the motion of a thin deflector to observable lensing quantities has
traditionally been regarded as too small to merit discussion even in
reference books in gravitational lensing. However, there have been
indications that the effect is actually linear in $v/c$.

The earliest indication that the leading correction is probably linear
in $v/c$ can be found in \citet{EFS}, where it is shown that the
internal motion of a deflector at rest affects the effective index of
refraction in first order (Eq. (4.16) of \citet{EFS}).  The effect is
considered negligible for the purposes of \citet{EFS}.

The next indication appears in \citet{pyne1}, where the authors develop
a method for integrating the null geodesics of a spacetime that is a
small perturbation of flat space.  As an application of the method,
the authors calculate the bending angle due to a point mass $m$,
obtaning

\begin{equation}
    \alpha=-\frac{4m}{r_0}(1-\bm{v\cdot n}_0)
\end{equation}

\noindent where $r_0$ is the distance of closest approach of the
lightray to the mass, and $\bm{n}_0$ is the direction of the null ray
at that point. The term $-\bm{v\cdot n}_0$ is the Doppler shift of the
deflector, $z_d$, so we can read this result as

\begin{equation}
    \alpha = (1+z_d)\alpha_s,
\end{equation}

\noindent which means that the bending angle is corrected by a factor
$(1+z_d)$ with respect to the bending angle of the same deflector at
rest in a Minkowski background. Here the quantity $z_d$ should not be 
confused with the cosmological redshift of the deflector.  

The third time the motion of the deflector was found interesting was
in \citet{italians}, the first of a series of papers where the leading
correction to the bending angle is calculated for different types of
deflectors. In \citet{italians}, by pursuing the integration of null
geodesics in a weak perturbation of flat spacetime by a slightly
different method, for a point mass the authors find

\begin{equation}
    \alpha = (1-2z_d)\alpha_s,
\end{equation}

\noindent [this is Eq.~(33) in \citet{italians}].

More recently, in \citet{fermatkling}, we use a completely different
method based on envelopes of null surfaces (as opposed to null
geodesics) to re-derive the lens equation, and we find  $\alpha =
(1+z_d)\alpha_s$ for the case of a lens moving along the line of
sight.

Aside from a purely academic interest in lensing by moving deflectors,
there has been a growing interest on the part of the microlensing sector of
practicing astrophysicists, stemming from improvements in observational
accuracy. Microlensing results when a background star in the Magellanic
Clouds (for instance) lines up with the line of sight between us and a
compact object in our Galaxy. The line up may happen due to the motion of
the star, or to the motion of the compact object, but the same treatment
is normally used in both instances. The motion creates a time dependent
brightness curve that peaks sharply when the star is closest to a caustic
created by the compact object in our past lightcone. The lightcurve is
usually studied using the static lens equation, assuming the source moves
with approximately uniform transverse velocity on the source plane.
Inverting the lens equation one obtains a curve for the motion of the
images on the lens plane which can be used to evaluate the magnification
as a function of time -- see for instance \citet{paczynski}.  The method
is used indistinctively for the case that the lens moves right through
the line of sight to a static source.  The point to be made is that there
are cases of moving lenses of interest, and corrections to the lens
equation of first order in $v/c$ (longitudinal speed) might become
observable as the technology keeps improving. An accurate
determination of the leading correction of order $v/c$ to the bending
angle of light rays due to the relative motion between the lens and the
observer may become necessary in the near future.

Here the calculation of the integration of the null geodesics in the case
of a time-dependent perturbation of a certain kind is pursued once more
with the methods and notation of \citet{EFS}.  Our method uses elements of
both of the calculations in \citet{pyne1} and \citet{italians}, as well as
new elements and thus functions as an independent cross-check. We find
full agreement with \citet{pyne1} and \citet{fermatkling}, and we back up
our calculation with a check of consistency with the leading correction to
the time delay, which, to our knowledge, has not appeared before except in
\citet{fermatkling}. The geodesic equation for linearized perturbations
corresponding to a non-stationary deflector of certain generality is
written down in Section~\ref{sec:2}, where the notation and approximations
are defined. The bending angle is evaluated in Section~\ref{sec:3}.  The
effects on a microlensing lightcurve are obtained in Section~\ref{sec:4}.
Concluding remarks are offered in Section~\ref{sec:5}.

%------------------------------------------------------------------
\section{Bending angle by a deflector in slow motion}\label{sec:2}

Although we are not interested in the most general linearized localized
perturbation off flat space, in this Section we write the equation for
the null geodesics in an asymptotically flat spacetime representing a
slowly moving lens of certain generality.  This is because most steps in
the derivation of the bending angle in the case of interest are more
general and would perhaps be useful in a different context.  Assuming
that the metric deviates only slightly  from a flat metric, we use
cartesian coordinates $x^a\equiv \{ t,\vec{x}\}$ (with $a=0,...,4$ and
$\vec{x}\equiv (x,y,z)$).  We assume that the metric
$ds^2=g_{ab}dx^adx^b=(\eta_{ab}+\gamma_{ab})dx^adx^b$ where
$\eta_{ab}=diag(1,-1,-1,-1)$ is given in
the following form

\begin{equation}\label{metric}
    ds^2 = (1+2\phi)c^2dt^2
           -8c\vec{V}\!\!\cdot\! d\vec{x}\,dt
        -(1-2\phi)d\vec{x}\!\cdot \!d\vec{x}
\end{equation}

\noindent where a dot in between two 3-dimensional quantities represents
the Euclidean scalar product ($\vec{u}\cdot \vec{w} \equiv \sum u^iw^i =
\sum u_iw_i$) and $\phi=\phi(\vec{x},t)$ and
$\vec{V}=\vec{V}(\vec{x},t)$. The meaning of $\phi$ or $\vec{V}$ is
irrelevant to the calculations that follow; only their size matters. We
assume that we have two independent smallness parameters $\epsilon$ and
$\delta$ such that $\phi \sim \delta$ and $\vec{V} \sim \epsilon\delta$.
The parameter $\epsilon$ represents, in some sense, the ratio of the
speed of motion of the deflector to the speed of light, and $\delta$
represents the strength of the Newtonian potential of the deflector's
mass density. All the calculations that follow assume that any integer
powers of $\delta$ or $\epsilon$ greater than 1 can be neglected, but
the the product $\epsilon\delta$ will be kept in view of the
independence of the two parameters. The reader should be aware that our
particular choice of metric and sizes is motivated by the case of a
perfect fluid in harmonic coordinates, in which the following
expressions for the potentials
\begin{eqnarray}
    \phi_f(\vec{x},t) &\equiv& -\frac{G}{c^2}\int
    \frac{\rho(\vec{x}',t)}{|\vec{x}-\vec{x}'|} d^3x'   ,\\
    \vec{V}_f(\vec{x},t) &\equiv& -\frac{G}{c^3}\int
    \frac{\rho(\vec{x}',t)\vec{v}(\vec{x}',t)}{|\vec{x}-\vec{x}'|} d^3x'
                            \label{Vgen}
\end{eqnarray}

\noindent hold in the near zone.  However, we are not interested in
these particular expressions for the potentials $\phi$ and $\vec{V}$,
and in fact we do not use them for any of the calculations in this
work. Any expressions for the potentials are allowed as long as they
lead to a metric that satisfies the Einstein equations up to terms of
order $\delta,\epsilon$ or $\epsilon\delta$. This includes the
potentials representing a lens that moves rigidly without changing
shape, which will be used in the next section, and obviously the
slowly-moving fluid as well.

We are interested in lightrays $x^a(s)$ with tangent vector $u^a=
(u^0,\vec{u})\equiv dx^a/ds$ traveling in this spacetime.  Because
$u^a$ is null, namely $u^au_a=0$, it is defined only up to overall
rescalings, and the scale factor can be adjusted with an appropriate
choice of parameter along the path. For convenience, we choose the
scaling so that $\vec{u}\cdot\vec{u} = 1$, namely: the space part of
the tangent has unit length in the Euclidean background. This is
equivalent to choosing to parametrize the lightray with its Euclidean
length $\ell$, in contrast to the wider practice of using affine
parametrizations for null paths. The Euclidean line element is
precisely defined by
\begin{equation}
    d\ell^2 = d\vec{x}\cdot d\vec{x},
\end{equation}

\noindent evaluated along the path. We thus have, by construction, a
lightray given by $x^a(\ell) = (t(\ell),\vec{x}(\ell))$ with tangent
\begin{eqnarray}
         u^0 &=& \frac{dt}{d\ell}, \\
    \vec{u}  &=& \frac{d\vec{x}}{d\ell}
          \equiv  \widehat{u},
\end{eqnarray}

\noindent where a hat $\widehat{\;}$ is used to denote a 3-vector
of unit length in the Euclidean space -- a directional vector.
With this choice of parametrization, the nullity condition
$u^au_a=0$ reads explicitly
\begin{equation}
    (1+2\phi)c^2(u^0)^2
           -8c\widehat{u}\!\cdot\! \vec{V}\,u^0
        -(1-2\phi)
    =0.
\end{equation}

\noindent Solving this equation for $u^0$ in a power series of
$\epsilon$ and $\delta$ of the form $u^0 = \overline{u}^0 + \delta
u^0_\delta + \epsilon u^0_\epsilon + \epsilon\delta
u^0_{\epsilon\delta}$ we find

\begin{equation}
    cu^0 = 1 - 2\left(\phi-2\widehat{u}\cdot\vec{V}/c\right).
\end{equation}

\noindent Thus the time component of the null tangent is found
readily from the nullity condition.  It is significant that the
deviation of $cu^0$ from zeroth order is of order $\delta$ at
least, but there is no deviation of order $\epsilon$.  Next, we wish to
obtain an equation for the spatial part of the null tangent,
which, because of our choice of parametrization, is reduced to the
vector $\widehat{u}$ which gives the direction of the null ray. We
obtain it from  the geodesic equation in the
following manner. The equation for geodesic curves with generic
parametrization (not necessarily affine) reads
\begin{equation}
    \frac{du^a}{d\ell}+\Gamma^a_{bc}u^bu^c -\mu u^a = 0,
\end{equation}

\noindent where the factor $\mu$ is free, reflecting the freedom in
the choice of parametrization (see, for instance, Eq.~(3.3.2) of
\citet{wald} and the subsequent discussion, or page 33 of
\citet{hawkingellis} as well). A vanishing $\mu$ gives
rise to an affine parametrization by definition. In our case,
however, $\mu$ is not zero, but is so far unspecified. We need only
consider the space components, namely

\begin{equation}\label{geo1}
    \frac{du^i}{d\ell}+ \Gamma^i_{00}(u^0)^2
              +2\Gamma^i_{0j}u^ju^0
              + \Gamma^i_{jk}u^ju^k
              - \mu u^i=0.
\end{equation}

\noindent For the connection coefficients of the linearized
metric $\Gamma^a_{bc}\equiv \frac12 \eta^{cd}(h_{ad,b}+h_{bd,a}-h_{ab,d})$, we
have
\begin{eqnarray}
    \Gamma^i_{00} &=& 4c\dot{V}^i+c^2\phi_{,i}  \\
    \Gamma^i_{0j} &=& 2c(V_{i,j}-V_{j,i})-\delta_{ij}\dot{\phi}\\
    \Gamma^i_{jk} &=& \delta_{jk}\phi_{,i}
             -\delta_{ik}\phi_{,j}
             -\delta_{ij}\phi_{,k}
\end{eqnarray}

\noindent where $_{,i}\equiv\partial/\partial x^i$ and $\dot{\;}\equiv
\partial/\partial t$. With these expressions for the connection coefficients,
Eq.~(\ref{geo1}) reads
\begin{eqnarray}\label{geo2}
    \frac{du^i}{d\ell}
    + (4c\dot{V}^i+c^2\phi_{,i}) (u^0)^2& &\nonumber\\
    + 2(2c(V_{i,j}-V_{j,i})-\delta_{ij}\dot{\phi})u^ju^0
    &&\nonumber\\
    + \Gamma^i_{jk}u^ju^k
    - \mu u^i&=&0.
\end{eqnarray}

\noindent Because $\Gamma^0_{ab}$ are all of order $\delta$ at
least, only the zeroth-order of $u^0$ is needed. Also notice that
$u^j(V_{i,j}-V_{j,i}) = -(\widehat{u}\wedge(\nabla\wedge
\vec{V}))_i$. Using these two facts, Eq.~(\ref{geo2}) becomes
\begin{eqnarray}\label{geo3}
    \frac{d\widehat{u}}{d\ell}+ \frac{4}{c}\dot{\vec{V}}+2\nabla\phi
        -4\widehat{u}\wedge(\nabla\wedge \vec{V})&&\nonumber\\
        -\left(\frac{2}{c}\dot{\phi}+2\widehat{u}\cdot\nabla\phi
              +\mu\right) \widehat{u}&=&0.
\end{eqnarray}

\noindent But since $\widehat{u}$ has unit length, then
$\widehat{u}\cdot d\widehat{u}/d\ell = 0$.  Taking the scalar product of
(\ref{geo3}) with $\widehat{u}$ we obtain

\begin{equation}\label{alpha}
\frac{4}{c}\widehat{u}\cdot\dot{\vec{V}}
        -\frac{2}{c}\dot{\phi}
              -\mu =0,
\end{equation}

\noindent which allows us to determine the unknown function $\mu$.
Using (\ref{alpha}) into (\ref{geo3}) we finally have
\begin{eqnarray}\label{geo5}
    \frac{d\widehat{u}}{d\ell}
    &=&
    -2\left( \nabla\phi-\widehat{u}\cdot\nabla\phi\widehat{u}
        -2\widehat{u}\wedge(\nabla\wedge
        \vec{V})\right)\nonumber\\
    && -\frac{4}{c}(\dot{\vec{V}}-\widehat{u}\cdot\dot{\vec{V}}\widehat{u}).
\end{eqnarray}

\noindent If for any vector $\vec{w}$ we define the components that are
perpendicular and parallel to the lightray $\widehat{u}$ by
$\vec{w}_{\perp}\equiv \vec{w}  -\widehat{u} (\widehat{u}\cdot\vec{w})$  and,
in the case of the gradient, $\nabla_{\perp} \equiv \nabla  -\widehat{u}
(\widehat{u}\cdot\nabla)$, this equation reads

\begin{equation}\label{geo6}
    \frac{d\widehat{u}}{d\ell}
    =
    -2\left( \; \nabla_{\!\!\perp}\phi
        -2\widehat{u}\wedge(\nabla\wedge \vec{V})
    +\frac{2}{c}\dot{\vec{V}}_{\!\!\perp}\right).
\end{equation}

Integrating $d\widehat{u}/d\ell$ along the path of the lightray from the
emission point to observation point gives the change in the direction of
the lightray's path at the observation point relative to the emitted
direction:

\begin{equation}
    \widehat{u}_{\tiny\mbox{out}}-
    \widehat{u}_{\tiny\mbox{in}}
    = \int_{\tiny\mbox{source}}^{\tiny\mbox{observer}}
        \frac{d\widehat{u}}{d\ell} (x^a(\ell))d\ell
\end{equation}

\noindent We now assume that the function $\phi$ is vanishing
outside of a region where it can accurately be described as the
Newtonian potential of a mass density localized in a region that
is small compared to the distance to the source and the observer.
In such conditions, the lightray will suffer a total deviation
that will be small compared with its original direction, and the
change will take place only within a small range of values of
$\ell$.  In such conditions, the argument $x^a(\ell)$ of the
integrand functions $\phi(x^a(\ell))$ and $\vec{V}(x^a(\ell))$ can
be taken as the original unperturbed path of the lightray, since
it deviates from such an unperturbed path only in first-order
quantities, but $\phi$ and $\vec{V}$ are already small. In this
context, then, the argument $x^a(\ell)=(t(\ell), \vec{x}(\ell))$
can be taken as
\begin{eqnarray}
    \vec{x} &=& \ell\widehat{u}_{\tiny\mbox{in}}
                   +\vec{x}_{\tiny\mbox{in}} \label{inpathx}\\
    t       &=& \frac{\ell}{c}+t_{\tiny\mbox{in}}\label{inpatht}
\end{eqnarray}

\noindent where $\widehat{u}_{\tiny\mbox{in}}$ is a constant unit vector, and
$(t_{\tiny\mbox{in}},\vec{x}_{\tiny\mbox{in}})$ is an arbitrary reference
point along the unperturbed path of the lightray. Typically, since the
deflector has a small size and is moving slowly compared to the speed of light,
it defines a localized neighborhood where
$(t_{\tiny\mbox{in}},\vec{x}_{\tiny\mbox{in}})$ may be thought to lie. In the
case of static deflectors,  $\vec{x}_{\tiny\mbox{in}}$ would be taken as the
impact parameter of the lightray with respect to a sensibly defined center of
the lens, whereas $t_{\tiny\mbox{in}}$ would denote the time of closest
approach of the lightray to the deflector.  In the general case,
$(t_{\tiny\mbox{in}},\vec{x}_{\tiny\mbox{in}})$ should be thought of as an
average spacetime location of the interaction between the lightray and the
deflector.

With these assumptions, we can define, as usual,  the bending angle
$\vec{\alpha}$ as the difference between the initial and final directions of
the lightray:

\begin{equation}
\vec{\alpha} \equiv \widehat{u}_{\tiny\mbox{in}}
        - \widehat{u}_{\tiny\mbox{out}}
\end{equation}

\noindent and we have an explicit formula

\begin{equation}\label{last}
    \vec{\alpha}
    =2\int_{\tiny\mbox{source}}^{\tiny\mbox{observer}}
        \left( \; \nabla_{\!\!\perp}\phi
        -2\widehat{u}_{\tiny\mbox{in}}\wedge(\nabla\wedge \vec{V})
    +\frac{2}{c}\dot{\vec{V}}_{\!\!\perp}\right)d\ell
\end{equation}

\noindent All $\widehat{u}$'s in the integrand, including those implicit in the
label $\perp$, take the value $\widehat{u}_{\tiny\mbox{in}}$, and the
quantities $\phi$ and $\vec{V}$ are evaluated at $x^a(\ell)$ as given by
(\ref{inpathx}-\ref{inpatht}).

It is interesting to notice that the motion of the deflector may
also affect the observed redshift of the source, as noted in
\citet{pyne1}. The wave vector of a photon traveling along the null
path is given by

\begin{equation}
    k_a \equiv (\omega,\vec{k}) = \frac{dx^a}{ds} ,
\end{equation}

\noindent if $s$ is an affine parameter along the null geodesic.
In our case, an affine parameter $s(\ell)$ must satisfy

\begin{equation}\label{affinesl}
\frac{d^2s}{d\ell^2} = \mu \frac{ds}{d\ell} ,
\end{equation}

\noindent where $\mu$ is small and is given by (\ref{alpha}).
Since $\frac{dx^0}{ds} = \frac{dx^0}{d\ell}/\frac{ds}{d\ell}$ and
since $\frac{dx^0}{d\ell} = u^0 = 1/c$ to zeroth order, then

\begin{equation}
    \lambda = \frac{2\pi}{\omega} = 2\pi c\frac{ds}{d\ell} ,
\end{equation}

\noindent and Eq.~(\ref{affinesl}) can be interpreted directly as
an equation for the wavelength of the traveling photon:

\begin{equation}\label{affinelambda}
\frac{d}{d\ell} \ln\lambda= \mu
\end{equation}

\noindent Since $\mu$ is small, this integrates to

\begin{equation}\label{redshiftraw}
   \frac{ \lambda_o - \lambda_e}{\lambda_e}
   \equiv \frac{\Delta\lambda}{\lambda}=-\frac{2}{c}
   \int_{\tiny\mbox{source}}^{\tiny\mbox{observer}}
    \hspace{-0.8cm}\dot{\phi} -
    2\widehat{u}_{\tiny\mbox{in}}\cdot\dot{\vec{V}}
    \; d\ell .
\end{equation}

%----------------------------------------------------------------------
\section{Application to a moving rigid lens}\label{sec:3}

The situation that we wish to examine is that of a deflector that is
moving without changing shape. We can obtain the metric of such a
deflector by applying a coordinate transformation to the metric of the
same deflector at rest. The metric of any deflector at rest in
coordinates $(\vec{x}',t')$ is entirely
determined by its Newtonian potential $\phi_s=\phi_s(\vec{x}')$ in the
form

\begin{equation}\label{metricstat}
    ds^2 = (1+2\phi_s)c^2dt'^2 -(1-2\phi_s)d\vec{x}'\cdot d\vec{x}'
\end{equation}

\noindent To say that the deflector is moving rigidly along a path
$\vec{\gamma}(t')$ is equivalent to making a change of coordinates
$\vec{x}'\to \vec{x} = \vec{x}'+\vec{\gamma}(t')$.  Assuming that the
acceleration of the path, $\ddot{\gamma}$ is of order quadratic in the
velocity at least (namely, much smaller than the current order of
approximation), we can accompany the
shift in spatial positions with a change in the time coordinate $t'\to t$
such that $dt = dt' + \dot{\vec{\gamma}}(t')\cdot d\vec{x}'/c^2$.  The
metric takes the form (\ref{metric}) with
\begin{eqnarray}
      \phi(\vec{x},t)
  &=& \phi_s(\vec{x}-\vec{\gamma}(t)),\label{phimoving}\\
   \vec{V}(\vec{x},t)
  &=& \frac{\vec{v}}{c}\;\phi(\vec{x},t),       \label{Vrigid}
\end{eqnarray}

\noindent where we are using the notation  $\dot{\vec{\gamma}}(t) \equiv
\vec{v}$. [The reader should be aware that if the acceleration is of
linear order in $v/c$, one can still ``fix'' a transformation for the
time coordinate, but in that case the metric will not take the form
(\ref{metric}), and the calculation of the bending angle must be done
over from the beginning.] Using these expressions we have
\begin{equation}\label{Vterm1}
     \widehat{u}_{\tiny\mbox{in}}\wedge(\nabla\wedge \vec{V})
   = \frac{\widehat{u}_{\tiny\mbox{in}}\!\cdot\! \vec{v}}{c}
     \nabla_{\!\!\perp}\phi
    -\widehat{u}_{\tiny\mbox{in}}\!\cdot\!\nabla\phi
     \frac{\vec{v}_{\!\perp}}{c},
\end{equation}

\noindent and also

\begin{equation}\label{Vterm2}
    \dot{\vec{V}}_{\!\!\perp}
  = \dot{\phi}\, \frac{\vec{v}_{\!\perp}}{c}
\end{equation}

\noindent where contributions of the form
$\phi\ddot{\vec{\gamma}}(t)$ have been neglected.  With (\ref{Vterm1})
and (\ref{Vterm2}), Eq.~(\ref{geo6}) becomes

\begin{equation}\label{geo7}
    \frac{d\widehat{u}}{d\ell}
    =
    -2\left( (1-2\widehat{u}\cdot\vec{v}/c)\nabla_{\!\!\perp}\phi
    +2\frac{\vec{v}_{\!\!\perp}}{c}\frac{d\phi}{d\ell}\right).
\end{equation}

\noindent where $d\phi/d\ell = \dot{\phi}/c +\widehat{u}\cdot\nabla\phi$.
Integrating Eq.~(\ref{geo7}) yields the bending angle:
\begin{equation}\label{geo8}
    \vec{\alpha}
    =2\int_{\tiny\mbox{source}}^{\tiny\mbox{observer}}
    \hspace{-0.8cm}(1-2\widehat{u}\cdot\vec{v}/c)\nabla_{\!\!\perp}\phi
    +2\frac{\vec{v}_{\!\!\perp}}{c}\frac{d\phi}{d\ell} \,d\ell.
\end{equation}

In physically reasonable situations we expect the deflector to be both
thin and weak, in the sense that the actual path of light $\widehat{u}$
deviates at most in first order from $\widehat{u}_{\tiny\mbox{in}}$
during the transit time through the deflector, which, under such
conditions, is also small. During the transit time the velocities are
approximately constant and can be pulled out of the integral sign.  As a
consequence, the second term in the right-hand side of Eq.~(\ref{geo8})
is proportional to the difference of the values of the potential $\phi$
at the source and observer. Both values vanish, with the consequence that
the second term does not contribute to the bending angle. The leading
contribution to the bending angle (for small transverse accelerations, as
assumed in the beginning) thus results from the first term in the
right-hand side of Eq.~(\ref{geo8}):

\begin{equation}\label{alphamoving}
\vec{\alpha}
    =2(1-2\widehat{u}_{\tiny\mbox{in}}\cdot\vec{v}/c)
    \int_{\tiny\mbox{source}}^{\tiny\mbox{observer}}
      \hspace{-0.8cm}  \nabla_{\!\!\perp}\phi(\vec{x},t)    \; d\ell,
\end{equation}

\noindent where $\phi$ is a function of $\ell$ through
(\ref{phimoving}) where $\vec{x}$ and $t$ are given by
(\ref{inpathx}-\ref{inpatht}).  For comparison, the bending angle of the same
deflector at rest is

\begin{equation}
\vec{\alpha}_s
    =2\int_{\tiny\mbox{source}}^{\tiny\mbox{observer}}
        \hspace{-0.8cm}\nabla_{\!\!\perp}
            \phi_s(\vec{x})        \; d\ell,
\end{equation}

\noindent with $\vec{x}$ given by (\ref{inpathx}). The
relationship between the two can be found, in this case, by
defining a new integration variable for (\ref{alphamoving}) which
we here denote by $q$:

\begin{equation}\label{q}
    q \equiv \ell
    -\widehat{u}_{\tiny\mbox{in}}\!\!\cdot\vec{\gamma}(t(\ell))
\end{equation}

\noindent In terms of this variable we have

\begin{equation}
      \vec{x}(\ell)-\vec{\gamma}(t(\ell))
    = q \widehat{u}_{\tiny\mbox{in}}
     + \vec{x}_{\tiny\mbox{in}}
     - \vec{\gamma}_\perp,
\end{equation}

\noindent where $\vec{\gamma}_\perp$ is essentially a constant,
as is $\vec{x}_{\tiny\mbox{in}}$. The differential form of
Eq.~(\ref{q}) is

\begin{equation}\label{dq}
            dq
    = ( 1 -\widehat{u}_{\tiny\mbox{in}}\!\!\cdot\vec{v}/c)\, d\ell.
\end{equation}

\noindent With this change of variable, Eq. (\ref{alphamoving})
becomes

\begin{equation}
       \vec{\alpha}
    = (1 -\widehat{u}_{\tiny\mbox{in}}\!\!\cdot\vec{v}/c)\,\vec{\alpha}_s.
\end{equation}

\noindent In terms of the Doppler shift of the deflector $z_d$,
since $\widehat{u}_{\tiny\mbox{in}}$ points towards the observer,
we have $\widehat{u}_{\tiny\mbox{in}}\cdot\vec{v}/c = -z_d$ so the
bending angle by a moving deflector in terms of the same deflector
at rest is

\begin{equation}\label{alphaspeed}
       \vec{\alpha}
    = (1+z_d)\, \vec{\alpha}_s.
\end{equation}

\noindent This result is in agreement with \citet{fermatkling}, where a
significantly different method was used, and the deflector was assumed
not to move across the line of sight.  An important point to emphasize is the fact that the
transverse motion of the deflector has no effect on the bending angle, as
long as it is approximately uniform, as noted by \citet{pyne1}.

A consistency check for the validity of (\ref{alphaspeed}) has already
been suggested by \citet{pyne1}: in the limit in which the thickness of
the deflector vanishes, this equation verifies the aberration due to a
Lorentz transformation between the frames in which the (absolutely thin)
deflector is at rest or in motion with speed $v$.

There is also a nonvanishing leading correction to the time delay that a
lightray suffers in the presence of a deflector in motion.  In order to
calculate it, we set $ds^2=0$ and solve for $dt$ from
Eq.~(\ref{metric}). In the current order of approximation this yields
\begin{equation}
    cdt=\left(1-2\phi+\frac{4}{c}\widehat{u}\cdot\vec{V}\right)d\ell.
\end{equation}

\noindent which, in the particular case of the deflector in rigid motion
becomes
\begin{equation}
    cdt=\left(1-2\phi
\left(1-2\frac{\widehat{u}\cdot\vec{v}}{c}\right)\right)d\ell.
\end{equation}

\noindent Integrating along the path of the lightray, we have the total
travel time
\begin{equation}
    ct=L -2\int_{\tiny\mbox{source}}^{\tiny\mbox{observer}}
\left(1-2\frac{\widehat{u}\cdot\vec{v}}{c}\right)\phi d\ell.
\end{equation}

\noindent where $L$ is the Euclidean length of the path. The gravitational
time delay is thus
\begin{equation}
    c\Delta t^g=-2\int_{\tiny\mbox{source}}^{\tiny\mbox{observer}}
\left(1-2\widehat{u}_{\tiny\mbox{in}}\!\!\cdot\vec{v}/c\right)\phi d\ell.
\end{equation}

\noindent We can evaluate the integral using the same methods as
in the case of the bending angle, and the same approximations: the
transit time is small, so the velocities can be
pulled out of the integral sign, and the same change of variable
helps put the result in terms of quantities associated with the
same deflector at rest.  We find
\begin{equation}
    c\Delta t^g
=\left(1-\widehat{u}_{\tiny\mbox{in}}\!\!\cdot\vec{v}/c\right)
    \int_{\tiny\mbox{source}}^{\tiny\mbox{observer}}
        \hspace{-0.8cm}(-2\phi_s) dq.
\end{equation}

\noindent The quantity $\int(-2\phi_s) dq$ is the gravitational time delay
suffered by the lightray in the presence of the deflector at rest. So we
have
\begin{equation}
    \Delta t^g
=\left(1-\widehat{u}_{\tiny\mbox{in}}\!\!\cdot\vec{v}/c\right)
    \Delta t^g_s.
\end{equation}

\noindent This formula is consistent with our bending angle because it
implies the correct relationship between time delay and bending:

\begin{equation}
\alpha= \nabla_{\perp} \Delta t^g.
\end{equation}

\noindent

Finally, we can evaluate the redshift formula (\ref{redshiftraw}) in the
case of interest. In the first place, notice that
\begin{equation}\label{dotphiprelim}
    \dot{\phi} = -\vec{v}\cdot\nabla\phi_s
               = -\vec{v}_{\!\!\perp}\cdot\nabla\phi_s
                 -\widehat{u}_{\tiny\mbox{in}}\!\!\cdot\vec{v}
                 \;
                 \widehat{u}_{\tiny\mbox{in}}\!\!\cdot\nabla\phi_s.
\end{equation}

\noindent On the other hand, $\dot\phi =
c(d\phi/d\ell-\widehat{u}_{\tiny\mbox{in}}\!\!\cdot\nabla\phi_s)$.
From these two relationships, we obtain

\begin{equation}
(1-\widehat{u}_{\tiny\mbox{in}}\!\!\cdot\vec{v}/c)
\widehat{u}_{\tiny\mbox{in}}\!\!\cdot\nabla\phi_s =
\frac{d\phi}{d\ell}
+\frac{\vec{v}_{\!\!\perp}}{c}\cdot\nabla\phi_s
\end{equation}

\noindent and thus
$\widehat{u}_{\tiny\mbox{in}}\!\!\cdot\nabla\phi_s$ is equal to
$d\phi/d\ell$ up to zeroth order in $v/c$.  This fact can be used
into (\ref{dotphiprelim}) to obtain

\begin{equation}\label{dotphi}
    \dot{\phi} =  -\vec{v}_{\!\!\perp}\cdot\nabla\phi_s
                 -\widehat{u}_{\tiny\mbox{in}}\!\!\cdot\vec{v}
                 \;
                \frac{d\phi}{d\ell}.
\end{equation}

\noindent Also, from (\ref{Vrigid}) and (\ref{dotphi}) we have
\begin{equation}\label{dotVrigid}
\dot{\vec{V}} = 0
\end{equation}

\noindent to leading order. Using (\ref{dotphi}) and
(\ref{dotVrigid}) into Eq.~(\ref{redshiftraw}) we have

\begin{equation}
\frac{\Delta\lambda}{\lambda_e} =\frac{2}{c}
   \int_{\tiny\mbox{source}}^{\tiny\mbox{observer}}
    \hspace{-0.5cm}\vec{v}_{\!\!\perp}\cdot\nabla\phi_s
    +\widehat{u}_{\tiny\mbox{in}}\cdot\vec{v}\frac{d\phi}{d\ell}
    \; d\ell .
\end{equation}

\noindent The integrals can be evaluated using the same methods as in the
case of the bending angle and time delay, and the same approximations.
The reader will have no difficulty obtaining the following

\begin{equation}
\frac{\Delta\lambda}{\lambda_e} =
\frac{\vec{v}_{\!\!\perp}}{c}\cdot\vec{\alpha}_s.
\end{equation}

\noindent The source will thus appear redshifted due to the speed of
the deflector across the line of sight. This agrees with \citet{pyne1}.

%----------------------------------------------------------------------------
\section{Microlensing effects}\label{sec:4}

The observation of image separation by a lens that is moving along the line of
sight will necessarily be affected by the motion of the lens.  However, since
the effect amounts to an overall factor, it may not be separable from the mass
of the lens, usually not known with sufficient certainty. This entails an
inherent error (which could be as large as 10\%) in the determination of the
mass of the deflector on the basis of the separation of multiple images -- in
particular the size of an Einstein ring--, independently of the accuracy of the
observations.

Additionally, since the bending angle affects not only the location but also
the brightness of the images, we can straightforwardly trace corrections to the
magnification equation in the general case, and in particular to the
lightcurves of microlensing events, in which case the images may not be
observationally separated but their combined brightness is regularly observed
as a function of time.

The theory of microlensing lightcurves as put forth by \citet{paczynski}
calculates the combined magnification of the two images of a lightsource by a
point mass:

\begin{equation}
A = A_1+A_2 = \frac{u^2+2}{u(u^2+4)^{1/2}},
\hspace{1cm}u\equiv \frac{r_0}{R_0},
\end{equation}

\noindent where $r_0$ is the location of the source projected on the lens plane
(namely, $r_0=\beta D_l$ in the notation of \citet{EFS}), and

\begin{equation}
    R_0^2 \equiv \frac{4GMD_lD_{ls}}{c^2 D_s}
         = r \alpha_s \frac{D_lD_{ls}}{D_s}
\end{equation}

\noindent where $r$ is the location of the image on the lens plane ($r=\theta
D_l$ in the notation of \citet{EFS}) and $\alpha_s=4GM/(c^2 r)$ is the bending
angle by a point mass at the origin of coordinates on the lens plane. The light
curve $\log A(t)$ is then obtained by setting $r_0 =
(d^2+v_{\perp}^2t^2)^{1/2}$ and letting the time flow, for a given impact
parameter $d$.  This corresponds exactly to the case where the lens is at rest
in the observer's frame of reference and the source moves across the lens plane
with a transverse velocity $v_\perp$. The lightcurve is applied to the opposite
case: that of a lens moving in the frame where both the source and observer are
at rest, under the assumption that the motion of the lens does not affect the
bending angle. What we have shown is that the application to this case is
accurate if the lens is moving transversally, but there is a correction if the
lens has also a component of the motion along the line of sight.

It is not complicated to calculate the correction to the lightcurve due to the
longitudinal motion of the lens.  This calculation amounts to the substitution
$R_0^2 \to (1+v_{||}/c)R_0^2$ in the lightcurve. A generic microlensing event
will provide the timescale $t_0 = R_0/v_\perp$ of the event, where $R_0$ will
carry the factor $(1+v_{||}/(2c))$. Consequently, unless the transverse
velocity of the lens or its distance are known, its (corrected) mass
cannot be determined from the timescale $t_0$. Statistical analysis
cannot yield the deflector's mass to a better certainty than a factor of 3
(page 454 of \citet{paczreview}).  Therefore, an entangled correction to the
mass of the size of $v_{||}/c$ would not even seem relevant.  

Nevertheless, following up on \citet{paczreview}, there are at least two
potential cases in which the masses of individual lenses can be determined
unambiguously, because the distances and proper motions of the lenses can be
measured independently of the microlensing event. The first case is that of
stars in globular clusters acting as lenses for the stellar background of
either the Galactic bulge, LMC or SMC \citep{pacz94}. The second case is that of
nearby stars with high proper motions acting as lenses for the more distant
stars in the Milky Way, LMC or SMC \citep{pacz95a}. Our analysis in this section
shows that in both cases, in addition to the proper motion and distance of the
lensing star, an unambiguous determination of the mass requires that the
redshift of the lensing star be measured, in order for the correction factor of
$(1+v_{||}/c)$ to be  determined. The mass could finally be obtained from the
timescale $t_0$ of the microlensing event. The relevance of the correction of
order  $v_{||}/c$ to the mass would depend on the accuracy of the measurements
of $t_0, v_\perp, v_{||}$ and the distance to the lensing star, all of which
can be expected to improve steadily in the future.

%----------------------------------------------------------------------------
\section{Concluding remarks}\label{sec:5}

Our leading correction to the bending angle by a point mass differs from
that obtained by \citet{italians} by a factor of -2. Both calculations use
very much the same method for the propagation of light and a metric
perturbation of the same form (borrowed directly from \citet{EFS}).  The
calculations depart from each other in the parametrization of the null
geodesics.  We use the Euclidean length of the path instead of the
affine length, which simplifies our calculations in great measure. Our
equation for the propagation of the lightray in the general case,
Eq.~(\ref{geo6}), differs from Eq.~(26) in \citet{italians} by the
presence of the time-derivative of the potential $\vec{V}$.  But because
the time-derivatives of the potentials do not contribute to the bending
angle of a point mass if the acceleration is small, the extra term is
negligible in the application of an almost uniformly moving point mass.
The major departure appears to occur in the integration of this equation
to yield the bending angle of the point mass.

On the other hand, our result for the point mass agrees with
\citet{pyne1}, a work that seems to have been overlooked, perhaps because
their method lies quite outside of the standard. In fact, Pyne \&
Birkinkshaw use a metric perturbation of flat space based on a
linearization of the Schwarzschild metric in Schwarzschild coordinates,
instead of harmonic coordinates. Their metric perturbation representing a
point mass contains off-diagonal terms, which are normally absent in the
standard representation of the metric of a stationary mass distribution
in the Newtonian limit, Eq.~(\ref{metricstat}). Additionally, their
calculation hinges strongly on the assumption that the Euclidean length
is an affine parameter of the null geodesics of the perturbed spacetime,
an assumption that is unwarranted in the intended order of approximation
since,  as implied by Eqs.~(\ref{affinesl}) and (\ref{alpha}), the affine
length differs from the Euclidean length in terms that are linear in the
strength of the Newtonian potential.  Thus some question as to the
validity of the  Pyne \& Birkinkshaw calculation of the leading order
correction in $v/c$ may have lingered, which we hope to have dispelled in
our current work.

Our work, in fact, extends the Pyne \& Birkinkshaw result to extended lenses of
any given deflection potential, being, thus, independent of the model.  This
agrees with the intuitive notion that the deflection caused by a distribution
of point masses moving with the same speed should be equal to the sum of the
deflections caused by the individual point masses in the distribution.
However, our result in this extended case departs quite sharply from that
derived by \citet{italians2}, where the bending angle due to the moving lens is
not proportional to the bending angle due to the same deflector at rest, but,
on the contrary, it is quite complicated and model-dependent.

\section*{acknowledgments}

I wish to thank Ted Newman for extremely valuable insights relevant to this
calculation, and for the pleasure of many enlightening conversations. I am
indebted to Olaf Wucknitz for pointing out an oversight in an earlier version
of this work. This work was supported by NSF under grant No. PHY-0070624 to
Duquesne University.

%\bibliography{references}   % run latex, bibtex, latex, latex
%\bibliographystyle{prsty}

\end{document}